 \documentclass[final,5p,times,twocolumn]{elsarticle}

\usepackage{amssymb}
 \usepackage{amsthm}

 \usepackage{lineno}

\usepackage{graphicx}
\usepackage{dcolumn}
\usepackage{bm}
\usepackage{hyperref}
\usepackage{amsmath}
\usepackage{amsfonts}
\usepackage{sidecap}
\usepackage{float}
\usepackage{parskip}
\usepackage{grffile}

\usepackage{color}
\usepackage{hhline}

\usepackage{xcolor,url}

\definecolor{vblue}{rgb}{0.19,0.20,0.56}

\hypersetup{
	bookmarks=true,         
	unicode=false,          
	pdftoolbar=true,        
	pdfmenubar=true,        
	pdffitwindow=false,     
	pdfstartview={FitH},    
	pdftitle={My title},    
	pdfauthor={Author},     
	pdfsubject={Subject},   
	pdfcreator={Creator},   
	pdfproducer={Producer}, 
	pdfkeywords={keyword1, key2, key3}, 
	pdfnewwindow=true,      
	colorlinks=true,       
	linkcolor=vblue,          
	citecolor=vblue,        
	filecolor=vblue,      
	urlcolor=vblue          
}

\makeatletter
\def\@eqnnum{{\normalsize \normalcolor (\theequation)}}
 \makeatother

\journal{Chaos, Solitons \& Fractals}

\begin{document}

\begin{frontmatter}



\title{Multifractal analysis of eigenvectors of small-world networks}


\author[label1]{Ankit Mishra}
\author[label2,label3] {Jayendra N. Bandyopadhyay}
\author[label1,label3] {Sarika Jalan}

\address[label1]{Complex Systems Lab, Discipline of Physics, Indian Institute of Technology Indore, Khandwa Road, Simrol, Indore-453552, India }
\address[label2]{Department of Physics, Birla Institute of Technology and Science, Pilani 333031, India}
\address[label3] {Center for Theoretical Physics of Complex Systems, Institute for Basic Science (IBS), Daejeon 34126, Republic of Korea}

\begin{abstract}
Many real-world complex systems have small-world topology characterized by the high clustering of nodes and short path lengths. It is well-known that higher clustering drives localization while shorter path length supports delocalization of the eigenvectors of networks. Using multifractals technique, we investigate localization properties of the eigenvectors of the adjacency matrices of small-world networks constructed using Watts-Strogatz algorithm.\ 
We find that  the central part of the eigenvalue spectrum is characterized by strong multifractality whereas the tail part of the spectrum have $D_{q}$ $\rightarrow$ $1$. 
Before the onset of  the small-world transition, an increase in the random connections leads to an enhancement in  the eigenvectors localization, whereas just after the onset, the eigenvectors show a gradual decrease in the localization. We  have verified an existence of sharp change in the correlation dimension at the localization-delocalization transition.
\end{abstract}

\begin{keyword}
multifractals, eigenvectors, localization, small-world network
\end{keyword}

\end{frontmatter}

\section{Introduction}
Since the pioneering work of Anderson on localization of electronic wave function in  disordered media, eigenvector localization has become a fascinating and  an active area of research \cite{anderson}.\ 
In his original  paper, Anderson argued that disorder introduced in the diagonal elements of a Hamiltonian matrix will lead to localization of the electronic
wave function.\ Later this theory successfully explained the phenomenon of metal-insulator transition.\
The theory of Anderson localization found its application in almost all the areas of physics including condensed matter physics \cite{cond_matter},
chaos \cite{chaos}, photonics \cite{photonics},  etc.\ Additionally, the phenomenon of localization-delocalization transition  has been investigated for different systems 
such as random banded matrix \cite{RBM}, power-law random banded matrix (PRBM) \cite{PLBM}, vibration in glasses \cite{vibration_glasses}, percolation systems \cite{percoln_threshold},  etc.\
Most of these studies concentrated in analyzing an impact of the diagonal and off-diagonal disorder on the localization properties.\  Further, there exist systems in which disorder  is originated from randomness in   their geometry, leading to extensive research on localization 
in topologically disorder systems 
\cite {topolo_disorder, topology_andr_insul, Anderson_topological}.\

Many complex systems can be described as graphs or networks consisting of {\it nodes} and {\it links}.\  The {\it nodes} correspond to the elements of a system and links 
represent the interactions between these elements.\ Various network models have been  proposed to capture and mimic properties of real-world complex systems,
among which Erd\"os-Renyi random network  \cite{random}, scale-free network \cite{albert}, and small-world network \cite{small-world} models have been the most popular ones.\  The small-world networks are characterized by high clustering coefficient and small characteristic path length arising due to the topological disorder or random distributions of the connections in an originally regular network.\
The real-world systems exhibiting topological disorder are ramified fractals, percolation networks, polymers \cite{Ramified_fractals, percolation, polymers},  etc.\ 
Other examples of real-world complex systems depicting the small-world characteristics include brain network \cite{brain} and ecological network \cite{food_web}.\ 
Here, we construct small-world networks using the Watts and Strogatz algorithm \cite{small-world} as follows.\ Starting from a regular network where each node is connected with its $k$ nearest neighbors, the connections are rewired randomly with a probability  $p_{r}$.\  For the intermediate rewiring probability, the network undergoes the small-world transition characterized by high clustering and low path-length.\ 

Further,  several dynamical processes on networks can be better understood by spectra of the corresponding adjacency and Laplacian matrices.\ For example, the small-world network as a quantum model has been studied in terms of localization-delocalization transition of the spectra of underlying adjacency matrices \cite{Locn_small_world_t1_t2}.\
Using the level statistics, it was shown that the small-world networks having diagonal disorder and rewired links  with different values of coupling constant manifest localization-delocalization transition at a critical rewiring probability \cite{Locn_small_world_t1_t2} .\ 
Furthermore, quantum diffusion of a particle localized at an initial site on small-world networks  was demonstrated to have  its diffusion time being associated with the participation ratio  and  higher for the case of regular networks than that of the networks with the  shorter path length \cite{diffusion}.\  
Further, quantum transport modeled by continuous-time quantum walk (CTQW) in  small-world networks has also been  investigated \cite{CTQW};  however,  here the small-world model was a bit
different than the one proposed by Watts and Strogatz in  \cite{CTQW}; additional bonds were added to a ring lattice to make it a small-world network.\ 
It was argued that adding a large number of bonds leads to suppression of the transition probability of CTQW which is just opposite to its classical counterpart, i.e., continuous-time random walk
where adding shortcuts leads to  an enhancement of the transition probability \cite{CTQW}.\ 

This  paper investigates localization properties of eigenvectors of the adjacency matrix of the small-world networks due to presence of disorder in the network's topology arising due to random rewiring of the links to the originally regular network structure.\
We emphasize  that
unlike the original Anderson tight-binding model having diagonal disorder and nearest-neighbor interactions, we do not introduce diagonal disorder and  rather consider long-range hooping (interactions).\
Using multifractal analysis, we analyze the localization properties of the eigenvectors of the adjacency matrix considering the entire eigenvalue
spectrum as the network undergoes topological transitions from an initial regular structure to a random structure via the small-world network  as a consequence of the links rewiring.\  We probe the localization properties of the entire eigenvalue spectrum since existence of even a fraction of delocalized eigenvectors has been shown to impart crucial changes in the behaviour of the corresponding system. For example, an infinitesimal fraction of delocalized eigenvectors have been shown to impact the transport properties of the underlying system \cite {CTQW} \cite{N_extend}.\
The idea of the multifractal system was first introduced by Mandelbrot \cite{manderbolt} which later found its application in various different areas of real world complex systems such as stock market data \cite{stock}, 
foreign exchange data \cite{foreign_exchange}, time-series data of sunspots \cite{time} traffic  \cite{ traffic} air-pollution  \cite{air_polution}, heartbeat dynamics, \cite{heartbeat},
etc.\
We find that for small values of the rewiring probabilities 
($p_{r}$ $\leq 0.01$), an increase in  
the $p_{r}$ values,  i.e., by increasing the randomness, leads to  an increase  in the degree of localization;  whereas for high rewiring probabilities
($p_{r}$ $\geq 0.01$), localization get decreased with an increase in the value of $p_{r}$.\  
Furthermore, we discover that it requires a very few number of rewiring, i.e. a very small amount of  deviation from the regular structure,
for the occurrence of the delocalization-localization transition of the eigenvectors captured using  IPR statistics.\ The probability density function of the logarithmic of IPR  shows  scale-invariance at the critical rewiring probability corresponding to the transition.\

\section{Method}  
A network denoted by {\it G} = \{V,\,E\} consists of set of {\it nodes} and interaction {\it links}.\ The set of {\it nodes} are represented by V = $\{v_{1}, v_{2}, v_{3}, $\ldots$, v_{N}\}$ and 
{\it links}
by E = $\{e_{1}, e_{2}, e_{3}, $\ldots$ ,e_{M}\}$ where  $N$  and  $M$  are size of $V$ and $E$ respectively.\ Mathematically, a network can be represented by its adjacency matrix $A$ whose elements 
are defined
as  $A_{ij}$ = $1$ if node $i$ and $j$ are connected and $0$ otherwise.\ Further, here we consider simple network without any self-loop  or multiple connections.\
The eigenvalues of the adjacency matrix $A$ are denoted by $\left\{\lambda_{1}, \lambda_{2}, \lambda_{3},\ldots,\lambda_{N}\right\}$ where
$\lambda_{1} \geq \lambda_{2} \geq $\ldots$ \geq \lambda_{N}$ and 
the corresponding orthonormal eigenvectors
as $\left\{\bm{x}_{1}, \bm{x}_{2}, \bm{x}_{3}, \ldots, \bm{x}_{N}\right\}$.\ 
Starting with a regular network in which all the nodes have an equal degree, we rewire each edge of the network with a probability $p_r$.\ 
This  procedure of the rewiring allows to transform a regular network with $p_r = 0$, to a random 
network with $p_r = 1$.\ In the intermediate $p_{r}$ values, network manifests the small-world behavior which is quantified by a very high 
clustering coefficient and a very small average shortest path length \cite{small-world}. We would also like to divulge important topological properties of the network capturing various topological transitions upon links rewiring. First, the initial regular network ($p_{r} = 0$) has periodic boundary condition and each node is connected to its ($k/2$) nearest neighbours on each side of it. Let the shortest distance between any given pair of the nodes $i$ and $j$ be denoted by $r_{i,j}$ and thus the average shortest path length of the network would be $r$ = $ {\sum_{i \neq j} {r_{i,j}}}/ {N(N-1)}$. For $p_{r}$ = 0, $r$ scales like $r \sim N/2k$ which leads to its hausdroff dimension being equal to 1. The hausdroff dimension $d$ can be determined by the scaling of $r$ with the network size $N$,  defined as $r \sim N^{1/d}$. When the initial regular network is perturbed, for $p_{r}$ $<$ $0.01$, the networks have finite dimensions i.e $r$ grows as $r \sim N^{\gamma}$ where  $0$ $<$ $\gamma$ $<$ 1. For $p_{r}$ = 0.001, fitting $r \sim N^{\gamma}$ yields $\gamma$ $\approx$ $0.27$ and $d \approx 3.7$. Upon an increase in  the rewiring probability which leads to occurrence of the small-world transition for $p_{r}$ $\geq$ $0.01$, $r$ scales like $r \sim \ln N$ which makes the network having infinite dimension \cite{sc_mf}. 

We investigate the localization property of the eigenvectors as the
network undergoes from the regular structure to a random one.\  Localization of an eigenvector means that a few entries of the eigenvector have much higher values 
compared to the others.\ We quantify localization of the $\bm{x}_{j}$ eigenvectors
by measuring the inverse participation ratio (IPR) denoted as $Y_{\bm{x}_{j}}$.\ The IPR of an eigenvector $\bm{x}_{j}$ is defined as \cite{ipr} 
\begin{equation}
Y_{\bm{x}_{j}}  = \sum_{i=1}^N (x_{i})_j^{4},  
\label{ipr}
\end{equation}
where $(x_{i})_j$ is the $i^{th}$ component of the normalized eigenvectors $\bm{x}_{j}$ with $j$ $\in\left\{1,2,3 \ldots ,N\right\}$.\ The most delocalized eigenvector $\bm{x}_{j}$ will have
all its components equal, i.e., $(x_{i})_j = \frac{1}{\sqrt{N}}$,  with IPR value being $1/N$.\ Whereas, for the most localized eigenvector, only one component of the eigenvector will be non-zero, and the normalization condition of the eigenvectors ensures that the non-zero component should be equal to unity. Thus the value of IPR for the most 
localized eigenvector is equal to $1$.\ It is also worth noting that there may exist  fluctuations  in the IPR  values for a given state $\bm{x}_{j}$  for different
realizations of the
network  for a given rewiring 
probability.\ We report the results for the  ensemble average 
$Y_{\bm{x}_{j}}$ which we define as a sum of  IPR values over $\bm{x}_{j}$ lying in the range 
$\lambda<\lambda_{j}<\lambda+d\lambda$ divided by the number of such eigenvectors $NP(\lambda)d\lambda$ in this range, where $P(\lambda)$ is the probability distribution function (PDF) of $\lambda$.\ We now elaborate the averaging process for discrete  eigenvalue spectrum.\ Let $\lambda^{R}$ = \{$\lambda_{1}$,$\lambda_{2}$, $\ldots$ ,$\lambda_{N\times R}\}$ such that  
$\lambda_{1} \leq \lambda_{2} \leq $\ldots$ \leq \lambda_{N\times R}$  is a  set of eigenvalues of a network for all  $R$ random realizations where $N \times R$ is the size of $\lambda^{R}$.\ 
The corresponding eigenvector set of the $\lambda^{R}$ are denoted by {$\bm{x}^R$} $=$ $\left\{\bm{x}_{1}, \bm{x}_{2}, \bm{x}_{3}, \ldots, \bm{x}_{N\times R}\right\}$.
We then divide $\lambda^{R}$ for a given value of $d\lambda$ into further $m$ subsets where $m$ = $(\lambda_{N \times R} - \lambda_{1}) $/$ d\lambda$.\  
For each
$\lambda^{j}$ $\subset$ $\lambda^{R}$ and  the corresponding eigenvectors $\bm{x}^{j}$ $\subset$ {$\bm{x}^R$},  $\forall j = 1,2,$\ldots$ ,m $; 
$\lambda^{j}$ = $\{\lambda_{1}, \lambda_{2},$\ldots$,\lambda_{l^{j}}\}$ and corresponding eigenvector $\bm{x}^{j}$ = $\left\{\bm{x}_{1}, \bm{x}_{2}, \bm{x}_{3}, \ldots, \bm{x}_{l^j}\right\}$  where $l^{j}$ is  the size of $j^{th}$  subset such
that $\sum_{j=1}^m l^{j}$ = $N \times R$ with a constraint that $\lambda_{l^{j}} - \lambda_{1^{j}} \leq d\lambda$ for each subset, the corresponding set of  IPRs for $\bm{x}^{j}$ 
will be $\{Y_{\bm{x}_{1}},Y_{\bm{x}_{2}}, $\ldots$, Y_{\bm{x}_{l^j}}\}$.\ 
Hence, the average IPR ($Y_{\bm{x}_{j}}(\lambda)$) for each subset $\bm{x}^{j}$ can be calculated as
$ \frac {\sum_{i=1}^{l^{j}} Y_{\bm{x}_{i}}} 
{l^{j}}$ where $\lambda$ is central value for each subset i.e. $\lambda$ + $\frac{d\lambda}{2}$ = $\lambda_{l^{j}}$ and $\lambda$ - $\frac{d\lambda}{2}$ = $\lambda_{1^{j}}$. Here, we have taken, $N = 2000$, $m = 200$ and $R = 50$ for each rewiring probability. All the physical quantities follow the same averaging procedure in
this paper. 

\begin{figure}[t]
	\centering
	\includegraphics[width=.48\textwidth]{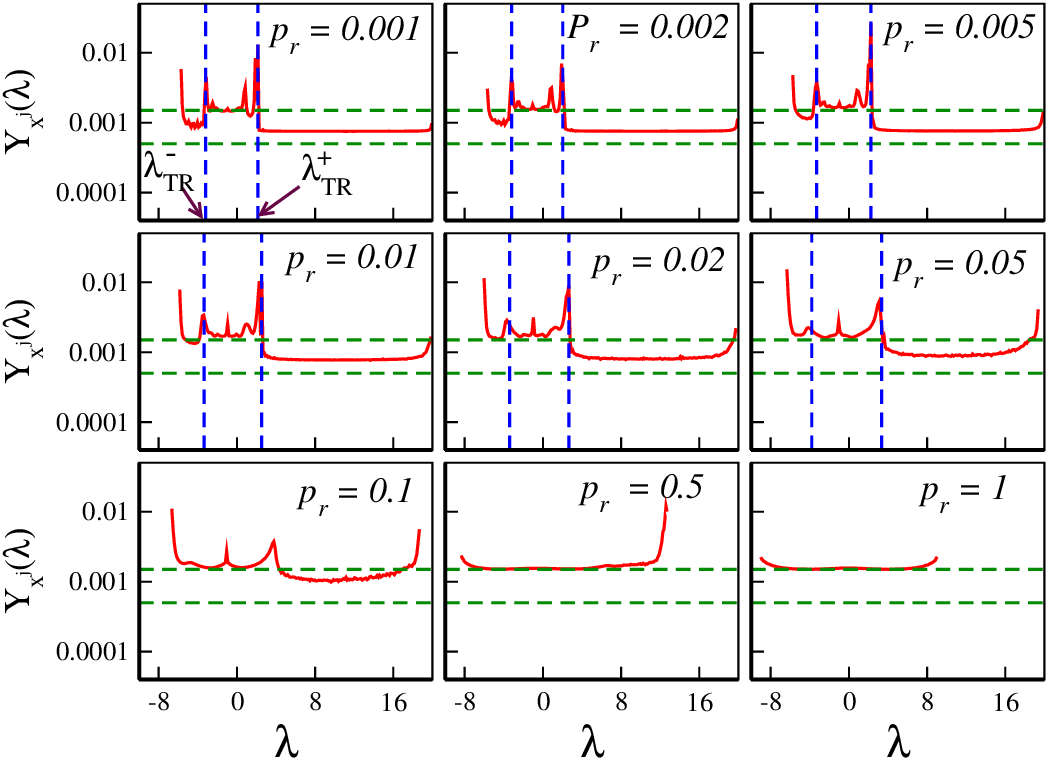} 
	\caption{IPR of the eigenvectors plotted as a function of the corresponding eigenvalues for various values of the rewiring probability.\
		The dashed green lines plotted at $0.0005$ and $0.0015$ correspond to the minimum possible value of the IPR ($1/N$) and the reandom matrix predicted value for the maximum delocalized state ($3/N$).\
		Here, $N = 2000$ and  $\langle k \rangle = 20$ are kept fixed for all the networks.}
	\label{ipr distribution}
\end{figure}
Further, in the seminal paper of Wegner \cite{Wegner} it was found that at the criticality, the generalized IPRs (GIPR) defined as $\chi_q = \sum_{i=1}^N {x_i}^{2q}$ shows
an anomalous scaling with the system size
$N$,  i.e., $\langle \chi_q\rangle \propto N^{-\tau(q)}$, where
$-\tau(q) = (q-1) \times D_q$.\ For the localized eigenvectors, $\langle \chi_q\rangle \propto N^{0}$, and for the completely delocalized eigenvectors $\langle \chi_q\rangle \propto N^{-d(q-1)}$ where d is the dimension of the system.\  However,
if the eigenvector corresponds to the critical state, $D_{q}$ becomes non-linear function of $q$ and therefore the scaling is described by many exponents $D_{q}$ indicating that  a critical eigenvector depicts multifractal behavior. 
We use the standard box-counting method as described in  \cite{cri_multifrac} for the multifractal analysis.\ 
Let us consider an eigenvector $\bm{x}_j$ whose components are represented as $(x_{1})_j, (x_{2})_j \ldots (x_{N})_j$.\ We then divide the $N$ sites into $N_L$ number of boxes 
with each box having the size $l$.\ The box probability $\mu^{k}(l)$ of the $k^{th}$ box of  the size $l$ is defined as
\begin{equation}
\mu^{k}(l) = \sum_{i= (k-1)l+1}^{kl} (x_{i})_j^2.
\label{multifractal}
\end{equation}
The $q^{\rm th}$ moment of the box probability is thus
\begin{equation}
\chi_q = \sum_k \mu_k^q(l) \sim l^{-\tau(q)}, 
\label{multifractal}
\end{equation}
\begin{figure} [t]
	\centering
	\includegraphics[width=.48\textwidth]{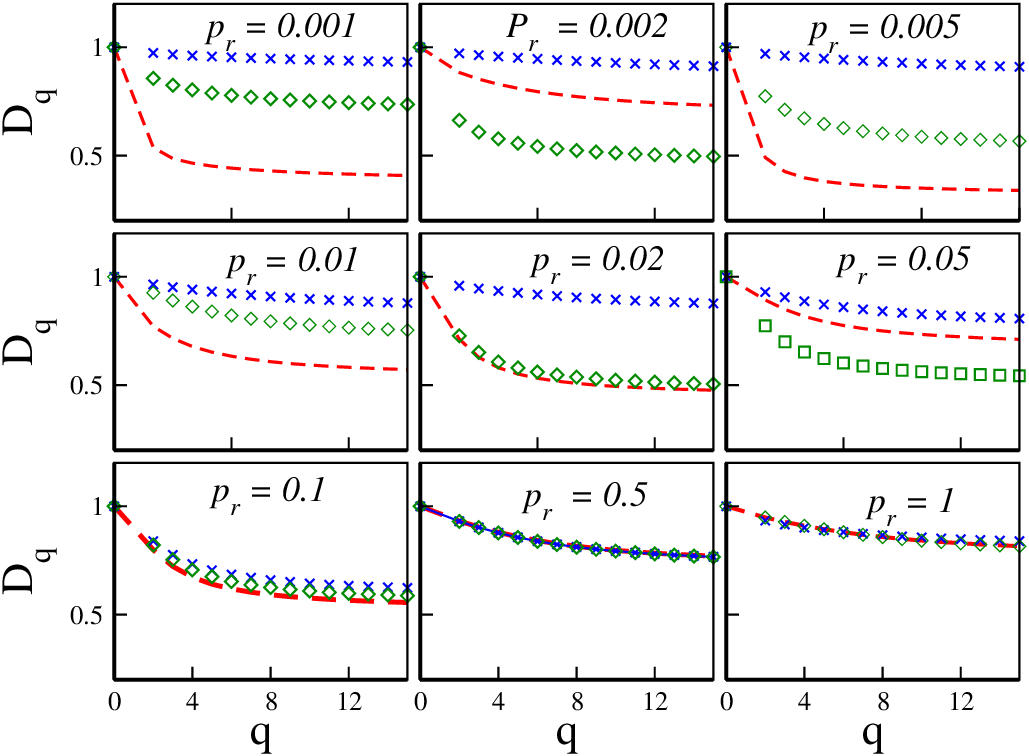} 
	\caption{The generalized fractal dimension $D_{q}$  plotted as a function of the exponent $q$ for the eigenvector 
		${\bm{x}^{j}}$ corresponding to $\lambda_{TR}^{+}, \lambda_{TR-1}^{+}, \lambda_{TR+1}^{+}$
		for various rewiring probabilities $p_r$. The $\color{red} ---$ , $ \color{blue} \times$,  $ \color{green} \diamond$  are for $\lambda_{TR}^{+}, \lambda_{TR+1}^{+}, \lambda_{TR-1}^{+}, $ respectively.} 
	\label{ev_multifractal}
\end{figure}

In the above equation, if the scaling exponent $\tau(q)$  is a linear function  of the parameter $q$, it corresponds to  the mono-fractal behavior, and for the nonlinear relation it indicates the multifractal property of the eigenvector.\ 

Note that, apart from the box counting method there exists an alternative method widely used in the localization theory through multifractal analysis. In this  method, instead of varying the length of the box ($l$), the system size ($N$) is varied by keeping the value of $l$ = 1 fixed.\ One usually first calculates  $\chi_q$ and  observes its scaling with the linear size of the system $L$ i.e. $\chi_q \sim L^{-\tau(q)}$.
This is also equivalent to $\chi_q$ $\sim$  $N^{-\delta(q)}$  where $N$ = $L^{d}$  and $\delta(q)$ = $\tau(q)$/$d$. The linear size of a network is defined as its diameter which is the longest of the shortest path between all the pairs of the nodes. Thus, approaching the problem through this method will require varying the network size  to a very large value which becomes computationally very exhaustive. We prefer box counting method in our analysis instead of the above described method. Nevertheless, both the methods will yield the same results. 

\section{Results} 
We analyze the localization properties of  the eigenvectors of adjacency matrix of networks with the variation of $p_{r}$.\ Rewiring of the connections  affects two major structural properties of the network: clustering coefficient ($CC$)
and average shortest path-length ($r$).\ 
For small values of the rewiring probability ($p_{r} < 0.01$), CC remains at very high value whereas r shows a drastic drop and attains a very low value. For $p_{r}$ $\geq$ $0.01$, there is no further change in $r$ as it has already attained a very low value but  $CC$ starts decreasing. Note that a higher clustering  is known to drive localization whereas a smaller path-length is believed to support delocalization. Thus,
to understand the contrasting impact of these two structural properties on the eigenvector localization, we first characterize the eigenvalue spectrum into different regimes based on the localization properties of the corresponding eigenvectors.
The central part ($\lambda_{TR}^{-} \leq \lambda \leq \lambda_{TR}^{+}$) of the spectrum  consists of critical eigenvectors having IPR of the order of
$10^{-3}$.\
This is the most localized part of the eigenvalue spectrum.\ Further,
$Y_{x_{j}}(\lambda)$ has U-shape for the smaller eigenvalues ($\lambda < \lambda_{TR}^{-} $) while it remains almost constant for the higher eigenvalues ($\lambda > \lambda_{TR}^{+} $) forming the tail part of the spectra.
The eigenvalues $\lambda_{TR}^{-}$ and $\lambda_{TR}^{+}$ separate the central part from the smaller and the larger eigenvalues, respectively.\
We refer to the central regime as a critical state regime, since here all the eigenvectors are at the critical state identified using the multifractal analysis.  Both sides of the central part are referred to as the mixed state as in this regime both the delocalized eigenvector (with IPR $\sim 10^{-3}$) and critical states eigenvector (with IPR $\sim 10^{-4}$) co-exist.
Using  the multifractal analysis, we first determine $\lambda_{TR}^{+}$ for various values of the rewiring probability.\ This will help us to know the nature of the change in
the width of the central part (critical states regime) with the change in the rewiring probability.\ We then analyze the change in  the degree of localization of eigenvectors with the change 
in the rewiring probability.\
\begin{figure} [t]
	\centering
	\includegraphics[width=.48\textwidth]{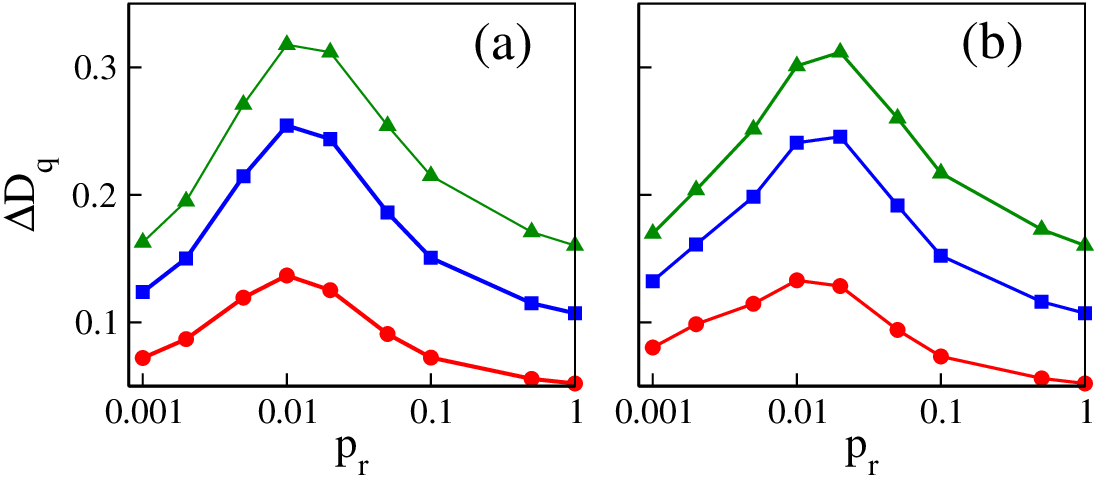} 
	\caption{Plot of $\Delta D_{q}$ as a function of rewiring probability $p_{r}$ for $q$ = $2\ (\color{red} \bullet) $ , $5\  (\color{blue} \blacksquare) $ and $10\ (\color{green} \blacktriangle) $ 
		respectively. (a) $\lambda \approx 1.271 $ (b) $\lambda \approx 1.371 $. These are the eigenvalues from the central regime.} 
	\label{delta_dq}
\end{figure}

{\bf Calculation of} $\bm\lambda_{\bm T \bm R}^{\bm +}$ $:$\  Figure \ref{ipr distribution} plots $Y_{x_{j}}(\lambda)$  as a function of $\lambda$ for various  values of the rewiring probabilities.\ 
All the different regimes, critical and the mixed can easily be identified from the Figure~\ref{ipr distribution} .\ 
First, we  discuss the impact of rewiring on the value of $\lambda_{TR}^{+}$, which  helps  us to  further  understand the change in
the width of the central part (critical states regime) with the variation in $p_{r}$.\ To achieve this, we analyze the multifractal behavior of a few 
eigenvectors  separating the critical regime with the mixed regime corresponding to the higher eigenvalue, i.e., $\bm{x}^j$  corresponding to $\lambda_{TR}^{+}$,  $\lambda_{TR+1}^{+}$, and  $\lambda_{TR-1}^{+}$.\ 
Here, $\lambda_{TR+1}^{+}$ and $\lambda_{TR-1}^{+}$ refer the eigenvalues just after and before $\lambda_{TR}^{+}$, respectively such that $\lambda_{TR-1}^{+}<\lambda_{TR}^{+}<\lambda_{TR+1}^{+}$ relation holds.
The eigenvectors $\bm{x}^j$  corresponding to $\lambda_{TR+1}^{+}$ are delocalized, hence we expect them having $D_{q}$ $\rightarrow$ $1$ ; whereas $\lambda_{TR-1}^{+}$  lies on the 
critical regime and therefore should have  a multifractal property.
Figure \ref{ev_multifractal} plots $D_{q}$ as a function of $q$ for the eigenvector $\bm{x}^j$  corresponding to $\lambda_{TR}^{+}, \lambda_{TR+1}^{+}$ and $\lambda_{TR-1}^{+}$ for different values of $p_r$.\
For $0.001 \leq p_{r} \leq 0.05$, $\bm{x}^j$ corresponding to $ \lambda_{TR}^{+}, \lambda_{TR-1}^{+}$ show the multifractal characteristics  accompanied by a wide range of the
generalized multifractals dimension values, on the otherhand, $\bm{x}^j$  corresponding to  $\lambda_{TR+1}^{+}$ have $D_{q}\rightarrow 1\, \forall$ $q>0$.\
We find that $\lambda_{TR}^{+}$ for various  values of the rewiring probability  lies in the range $2.03 \leq \lambda_{TR}^{+} \leq 3.34$,
i.e., there exists no significant change in the value of $\lambda_{TR}^{+}$ with the  change in the rewiring probability for fixed value of network parameters such as size $N$ and the average degree $k$ and hence width of the central part remains almost  fixed.\ 
We furthermore notice that $\lambda_{TR}^{+}$ for various  values of the rewiring probability always remains equal to the boundary of the bulk part of the
eigenvalue density ($\rho(\lambda$)) and the tail part of the eigenvalue density. Since the radius of the bulk part of the eigenvalues largely depends on the networks parameters \cite{camellia}, it is not surprising that the  value of $\lambda_{TR}^{+}$ remains almost same. The tail part of the eigenvalue spectrum have very low value of probability density. Mathematically, this means  that,
$\rho$($\lambda_{TR}^{+}+\epsilon$) $\rightarrow$ $0$ and $\rho$($\lambda_{TR}^{+}-\epsilon$) $\rightarrow$ $\delta$ where
$\delta > \epsilon$ and $\epsilon$ $\ll$ $1$.\ This can be easily understood with the following argument.\   The eigenvalue  spacing $\lambda_{i+1}$$-$$\lambda_{i}$ $\ll$ $\xi$ for 
$\lambda$ $<$ $\lambda_{TR}^{+}$  whereas for $\lambda$ $>$ $\lambda_{TR}^{+}$, $\lambda_{i+1}$$-$$\lambda_{i}$ $>$ $\zeta$ where $\zeta$ $>$ $\xi$ and $\xi$ $\ll$ $1$.\ 
It has been argued that the eigenvalue spectra corresponding to the localized states is continuous while that of  the  eigenvalue spectra of the delocalized states is discrete \cite{spec_loca_deloca}.\ Thus, $\lambda_{TR}^{+}$ is 
the eigenvalue separating the central regime of high IPR value with the regime of  the low  IPR values.\  Additionally, it also separates the bulk part from  the tail of the density of the eigenvalues.\ Therefore,  these 
calculations of eigenvalues separating the regime of higher IPR values with lower IPR values  are in agreement with the previously known  conjecture  on
localization.\ More interestingly,  the  methodology  followed here provides the exact value of the  eigenvalue separating the critical regime and mixed regime.\
After $p_{r} \geq 0.05$,  as randomness  increases further, it  is difficult to divide the spectrum into  different  regimes based on the localization properties of eigenvectors and all the three regimes start coinciding with each other (Figure \ref{ipr distribution}).\  Additionally, it can be seen that  $D_{q}$ for
$\bm{x}^j$  corresponding to  $\lambda_{TR}^{+}, \lambda_{TR+1}^{+}$ and $\lambda_{TR-1}^{+}$  starts coinciding with each other (Figure \ref{ev_multifractal}).  

\begin{figure} [t]
	\centering
	\includegraphics[width=.48\textwidth]{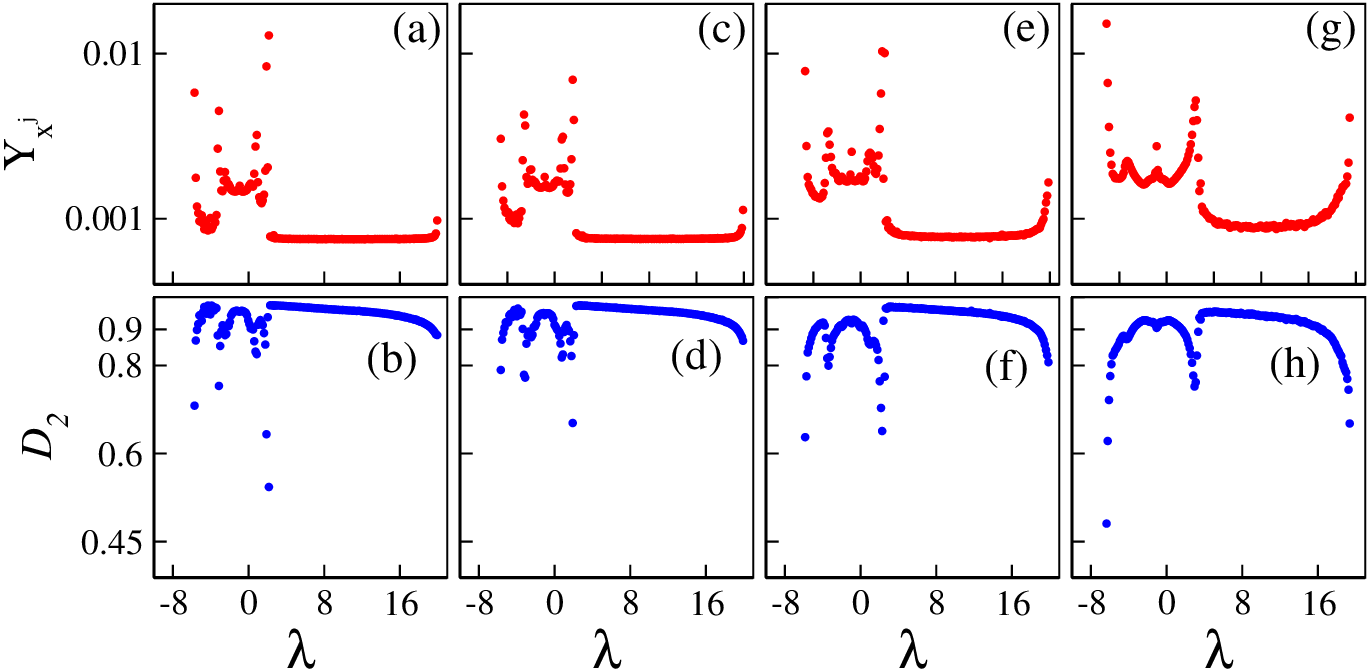} 
	\caption{The correlation dimension $D_2$ and IPRs of the eigenvectors are plotted as a function of the corresponding eigenvalues for the following four different rewiring probabilities: 
		(a)-(b) $p_r = 0.001$; (c)-(d) $p_r = 0.0021$; (e)-(f) $p_r = 0.01$; and (g)-(h) $p_r = 0.05$.}
	\label{D2}
\end{figure}

{\bf Change in the localization properties with  $p_{r}$}:  We next discuss the impact of rewiring on the degree of localization of the eigenvectors.\ Specifically, we focus on the 
eigenvectors belonging to the central  regime as this part of the spectrum undergoes the localization-delocalization transition with the increase in the rewiring probability.\
The other eigenvectors lying outside the central regime do not  witness significant  change in their localization properties.\ For $p_{r}$ = $0.001$, the eigenvectors which are nearer to the band edge, i.e. $\bm{x}^j$ corresponding to $\lambda_{TR}^{+}$, $\lambda_{TR-1}^{+}$  are characterized by strong multifractality having a wide range of the generalized multifractal dimensions.\ On the other hand, the eigenvectors inside the band are characterized by weak multifractality satisfying $D_{q}$ = $1-\beta q$ $\forall$ $q>0$  and $\beta \ll$ 1.\ The weak multifractality  means that the eigenvectors corresponding to the critical state are close towards the extended states which are analogous to Anderson transition in $d = 2+\epsilon$ with $\epsilon \ll$ 1 dimension.\ Furthermore,  a strong multifractality means that 
the corresponding eigenvectors is more inclined towards the localization which is similar to the conventional Anderson transition in $d \gg$ 1 dimensions \cite{Anders_trans}.\ 
As the rewiring probability is increased further, for  $0.001<p_{r} \leq 0.05$, we  do not find any significant change in the multifractal characteristics of  the
eigenvectors lying at the band edge.\ However,  the eigenvectors lying inside the band are now described by the strong multifractal characteristics.\ To demonstrate the change in the strength of multifractality of the eigenvectors 
with the variation in  $p_{r}$, 
we calculate the decay in $D_{q}$  with respect to $q$.\ For this, we define $\Delta D_{q} = D_{0}-D_{q}$.\ Note that, $D_{0}$ = $d$ (in our case: $d = 1$) irrespective of the nature of the eigenvector.
Hence, $\Delta D_{q}$ provides a correct measure to compare the decay in  $D_{q}$  with respect to $q$ inturn providing insight about the strength of multifractalty.\ Thus, we use $\Delta D_{q}$ as a measure of the degree of
localization.\ Figure \ref{delta_dq} plots $\Delta D_{q}$ as a function of $p_{r}$ for two different eigenvalues from the central part.\ 
Figure \ref{delta_dq} demonstrates that as  the rewiring probability increases, $\Delta D_{q}$  manifests an increase  until the onset of small-world transition ($p_{r} \approx 0.01$).
Thereafter, it shows a decrease for a further  increase in the $p_{r}$ values till $p_r = 1$ .\  The increase in $\Delta D_{q}$ for the initial $p_{r}$ values indicates that there exists an increase in the 
multifractal characteristics indicating an enhancement in the degree of localization with the increase in the rewiring probability.\
Thus, based on the effect of rewiring of the connections on the localization properties of the
eigenvectors, $p_{r}$ can be divided into two domains.\ First, $0.001<p_{r}\leq 0.01$  where  an increase in the rewiring probability leads to  an increase in the degree of localization of eigenvectors; 
while for $0.01 \leq p_{r} \leq 1$, eigenvectors undergo a continuous decrease in the localization.\ Moreover,
the transition takes place exactly at the onset of the small-world transition.\ This can be  further  explained  by  the following.\ For $0 < p_{r} \leq 0.01$,  the average clustering coefficient of the network remains 
constant at $CC = 3/4$, while the average shortest path drops down drastically.\ It is a common belief that a shorter $r$ support diffusion whereas a  higher clustering is known to
drive  toward the localization transition \cite{CC_locn}.\  Therefore, there  exists an interplay of these two  structural quantities on deciding the localization properties of the
eigenvectors.\ For  $p_{r} \leq 0.01$, a high number of triangles  accounted for localization, whereas for $p_{r}\geq 0.01$, there  exists a significant
decrease in the CC with $r$ being small,  and thereby leading to decrease in the degree of localization of eigenvectors.\ 
We would like to stress that distorting the initial regular network  by rewiring a few connections (for $p_r$ being very small) does not cause localization of the eigenvectors, rather they reach to the critical states detected by the  calculation correlation dimension ($D_{2}$).\ 

\begin{figure} [t]
	\centering
	\includegraphics[width=0.52\textwidth]{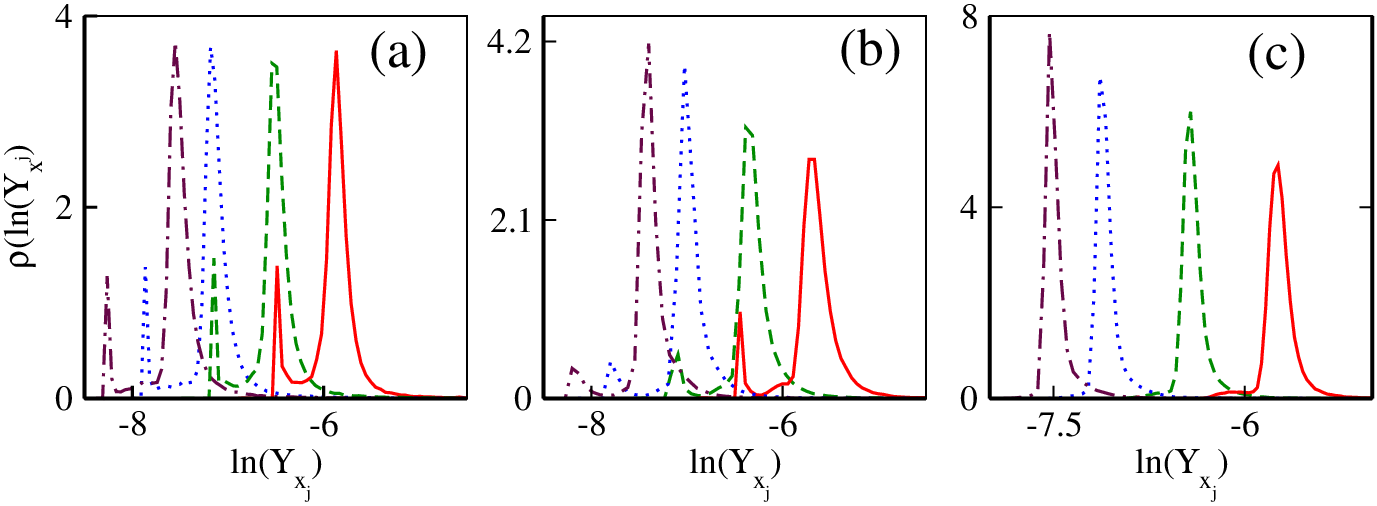} 
	\caption{ Distribution function of $P(ln(Y_{x_{j}}))$  for rewiring probability (a) 0.001 (b) 0.01 (c) 0.2.\ $\color{red} -$, $ \color{green} ---$,  
		$\color{blue}\cdots $, $ \color{brown} -\cdot-\cdot-$ are used for $N = 1000, 2000, 4000, 6000$ respectively. }
	\label{L_Ipr_N}
\end{figure}

The correlation dimension ($D_{2}$) of the eigenvectors provides insight into the scaling of the IPRs.\
For the localized eigenvectors, $D_{2}$ $\rightarrow$ $0$, while $D_{2}$ $\rightarrow$ $1$ for
the completely delocalized eigenvector.\ On the otherhand, if  $0< D_{2}<1$, the eigenvector is said to be at the critical state.\ Therefore, we next calculate the correlation dimension of the eigenvectors for various values of $p_{r}$.\ Figure \ref{D2} presents results of $D_2$ and IPR for the eigenvectors as a function of the corresponding eigenvalues for four different $p_{r}$ values.\ 
The plot clearly depicts that there exists  a sharp change in $D_{2}$ at a point which separates the central and the delocalized regime.\ For $\lambda$ $>$ $\lambda_{TR}^{+}$, $D_2$ $>$ $0.94$
for all  the value of the rewiring probability indicating delocalized eigenvector.\ However, at the critical point, which separates the critical and the mixed regimes, the values of $D_2$
is different for the different $p_r$ values.\ For $p_r = 0.001, 0.002, 0.01$ and $0.05$, the values of $D_2$ are equal to $0.53, 0.66, 0.66$  and  $0.72$, respectively.\ The range $0.4<D_{2}<0.90$ for the
eigenvectors in the critical regime clearly suggests that though they reach at the critical state arising due to the links rewiring, they do not get completely localized ($D_{2} \rightarrow 0$).\
Further, the value of $D_{2}$ at  $\lambda_{TR}^{+}$ is minimum for the entire spectrum.\ Thus,the eigenvectors at the boundary of the central part are  the most localized in the spectrum for the  values of the initial rewiring 
probabilities.\

{\bf IPR Statistics} : So far, we have discussed the impact of rewiring  on the localization properties of the eigenvectors when IPR and other physical measures  of eigenvectors are
being averaged over in the small eigenvalue window.\
\ However,  an  analysis of the IPR statistics 
can provide us further information about  the system.\ For instance, in the case of power-law random banded matrix (PRBM), it was found that at the 
critical point of  the localization-delocalization transition, the width and the shape of the distribution of the logarithm of IPR  do not change with the system size, or we can say that it is scale invariant \cite{ipr_N}.\  We  calculate the distribution function of IPR for various rewiring probabilities for different 
system sizes.\ Figure \ref{L_Ipr_N}  shows that, for $p_{r}$ = 0.001,  $\rho(\ln(Y_{x_{j}}))$ remains invariant  with  the change in the network size  as neither  its  shape  nor its width changes
with $N$.\ 
For $p_{r} \geq 0.001$
the distribution function $\rho(\ln(Y_{x_{j}}))$ witness a continuous decrease in the width  with an increase in $N$ (Figure \ref{L_Ipr_N}).\ Thus, we  can  infer
that, $p_{r}$ = $0.001$ is the critical
rewiring probability for the localization-delocalization transition.\

\begin{figure} [t]
	\centering
	\includegraphics[width=.48\textwidth]{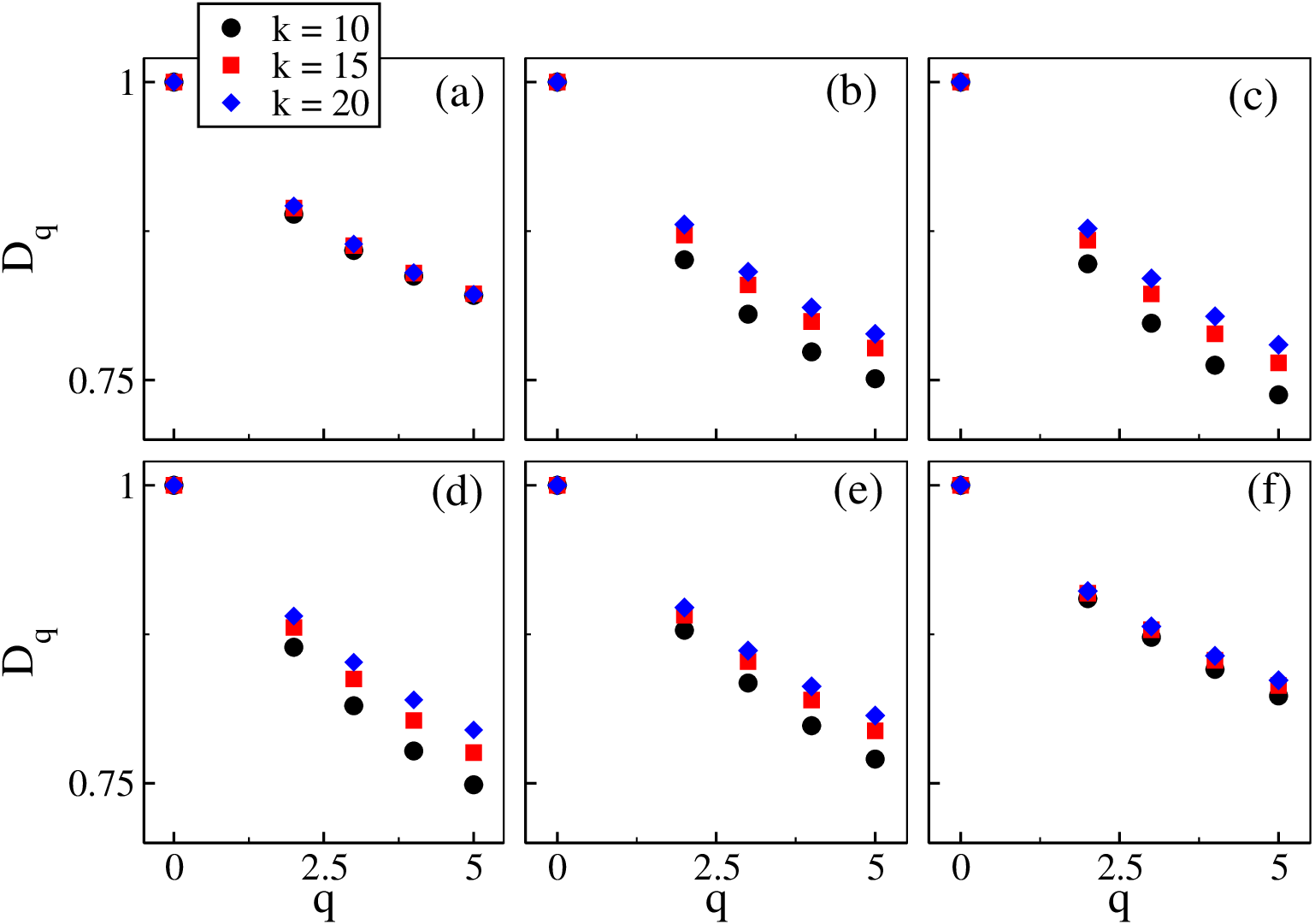} 
	\caption{Plot of $D_{q}$ as a function of q and $k$ for various rewiring probabilities. (a) $p_{r}$ = 0.001 (b) $p_{r}$ = 0.005 (c)
		 $p_{r}$ = 0.01 (d) $p_{r}$ = 0.05 (e) $p_{r}$ = 0.1 (f) $p_{r}$ = 1. In all the cases , $\lambda$ $\approx$ $0$ is considered and $D_{q}$ is average over all the eigenvectors belonging to d$\lambda$ $=$ $0.25$ as described in method section.  }
	\label{Dq_k}
\end{figure}

\begin{figure} [t]
	\centering
	\includegraphics[width=.48\textwidth]{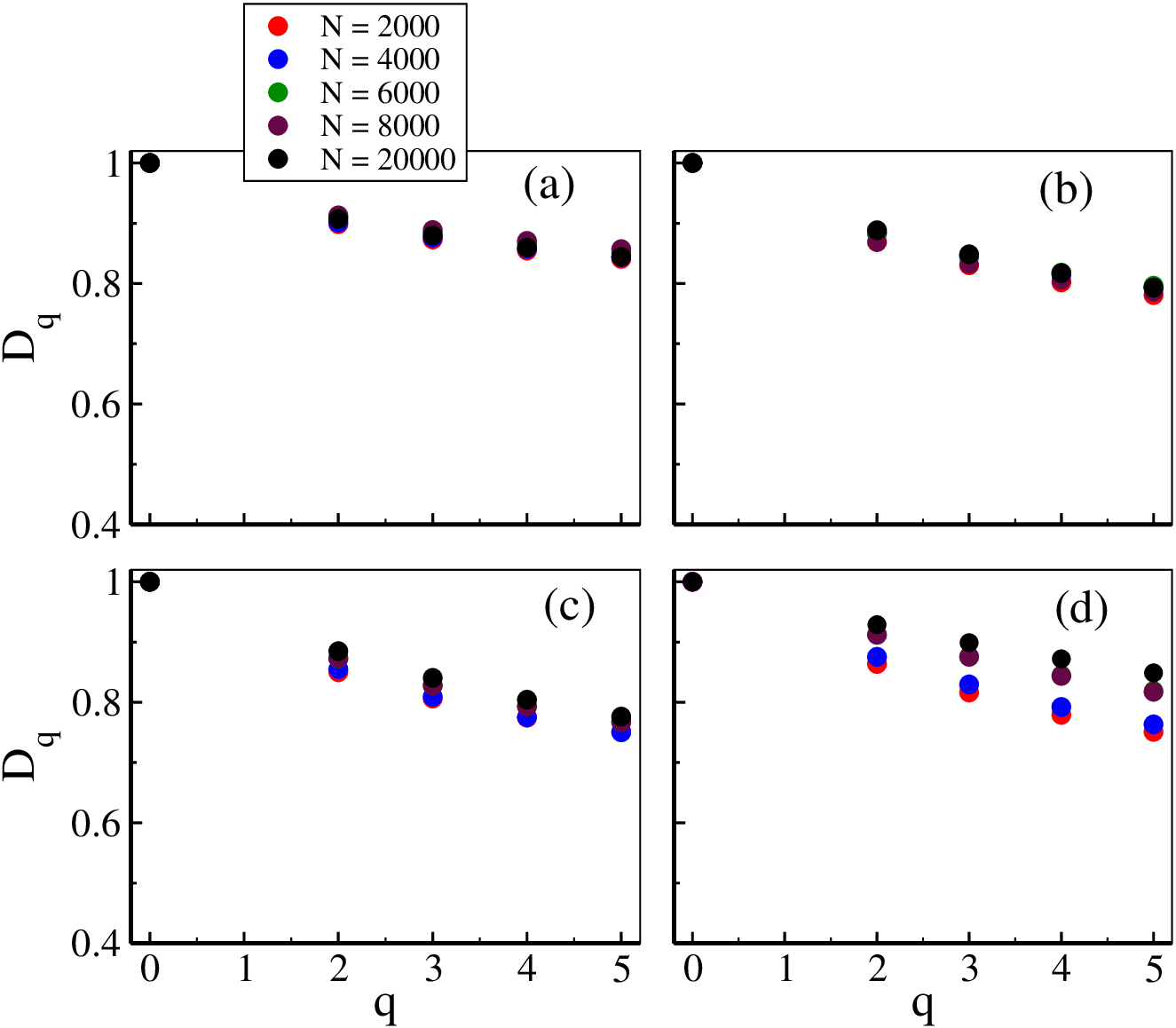} 
	\caption{Plot of $D_{q}$ as a function of q and $N$ for various rewiring probabilities. (a) $p_{r}$ = 0.001, $\lambda$ $\approx$ $1.47$ (b) $p_{r}$ = $0.005$, $\lambda$ $\approx$ 1.403 (c)
		 $p_{r}$ = $0.01$, $\lambda$ $\approx$ = $1.527$ (d) $p_{r}$ = $0.05$, $\lambda$ $\approx$ $1.69$. Here, $\lambda$ belongs to the central regime and $D_{q}$ is average over all the eigenvectors belonging to d$\lambda$  as described in Sec.Method.  }
	\label{Dq_N}
\end{figure}

\section{Impact of variation in average degree ($k$) on $D_{q}$}
In this section, we discuss the impact of average degree $k$ on $D_{q}$. Note that the largest eigenvalue of network is bounded with the largest degree $k^{max}$ \cite{camellia}.  Moreover, for a random network $\lambda_{1}$ $\approx$ $[1+o(1)]k$, where $o(1)$ means a function that converges to $0$ \cite{lam_bound}.  Thus, varying $k$ may affect the eigenvalue spectrum drastically even for the fixed network size. Here, we have considered three sets of $k$ = $10$, $15$, $20$ with $N$ = $2000$ being fixed. First we calculate the value of $\lambda_{TR}^{+}$ for $k$ = $10$ and $15$. For such a small change in the average degree though leads to notable changes in $\lambda_{1}$, there  exist no such significant impact of $k$  on $\lambda_{TR}^{+}$.

We next probe impact of $k$ on $D_{q}$ for various values of the rewiring probabilities. We witness no significant changes in the nature of $D_{q}$ for the tail part of the eigenvalue spectrum ($\lambda$ $>$  $\lambda_{TR}^{+}$) with the change in the average degree. However, for the central regime, a  decrease in the average degree leads to an increase in the strength of the multifractality of the eigenvectors as depicted in Figure \ref{Dq_k}.As we have already discussed that the strength of the multifractality of a eigenvectors indicates about the degree of localization. Thus, decreasing the average degree suggests  an enhancement in the degree of eigenvector localization.

\section{Effect of Finite Size}
It is well known that critical phenomenon accurately defined only in the thermodynamics limit i.e., $N$ $\rightarrow$ $\infty$. Further, multifractality of the eigenvectors might be due to the finite-size of the system, which may not exist at the infinite size limit. Hence, one needs to be careful regarding the critical point. Nevertheless, $D_{q}$ certainly reveals the tendency towards a more localized or a more delocalized behavior of a given eigenvector. Therefore, we have calculated $D_{q}$ for various system sizes to check the impact of finite-size effect in our analysis.
Figure \ref{Dq_N}, $D_{q}$ is ploted for the eigenvalues lying in the central regime for various different values of the rewiring probability as the network size is varied from $2000$ to $20000$.It is evident from the Figure \ref {Dq_N} (a,b) there is no significant change in $D_{q}$ as the network size is changed from $2000$ to $20000$. This is also supported by the Figure \ref{L_Ipr_N} (a), where the distribution  $\rho(\ln(Y_{x_{j}}))$ remains scale invariant and thus giving rise to the unique fractal dimension $D_{q}$. 
However, there exists a slight change between $D_{q}$ at $N =2000 $ and $D_{q}$ at  $N = 20000$
in the case of $p_{r}$ = $0.01$ [Figure \ref {Dq_N} (c) ] though $D_{q}$ gets saturated after  $N = 8000$. However,
we do find a significant change in $D_{q}$ with a change in the network size in the case of $p_{r}$ = $0.05$ though it still  keeps showing the multifractal characterstics.
Thus, we see that change in the value of $D_{q}$ by varying $N$ increases with an increase in the rewiring probability which appears very intriguing. One of the possible reasons for the larger fluctuations in  Figure \ref {Dq_N} (d) could be the higher rewiring probability as for a given rewiring probability, the number of the rewired links ($N_{r}$) on average equals to ($N \times p_{r} \times k$)/$2$. Thus, for Figure \ref {Dq_N} (d), $N_{r}$ equals to $10^{3}$ and $10^{4}$ for $N = 2000$ and $20000$, respectively. This difference is very high as compared with that of the smaller rewiring probabilities  leading to higher changes in the network topology with higher $N$.
Note that fluctuation of $D_{2}$ for a critical eigenvector of power-law random banded matrix (PRBM) with system size was also reported and investigated in \cite{Dq_N}. 

\section{ Discussion and Conclusion}
 We have investigated the localization behavior of the eigenvectors of the small-world networks.\ First, we characterize the eigenvalue spectrum into different regimes.\ The central regime corresponds to the critical state eigenvectors and the mixed regime where we found delocalized eigenvectors along with some critical states eigenvectors.\
Using the multifractal analysis, we find that there exists no significant change in the eigenvalue ($\lambda_{TR}^{+}$) separating the central regime and the mixed regime.\ Additionally, we notice no  significant change in
$\lambda_{TR}^{+}$ with an increase  in $N$, i.e. for $N \rightarrow \infty$, $\lambda_{TR}^{+}$($N$) $\sim$ $\mathcal{O}(1)$.\ Further, we demonstrated that the rewiring procedure can be divided into two domains.\
For  small rewiring, $p_{r}$ $\leq$ $0.01$, with an increase in the random connections, there exists a continuous enhancement in the localization of the eigenvectors corresponding to the central regime, while for 
the higher rewiring probability  $p_{r}$ $\geq$ $0.01$, eigenvectors gradually loose their degree of localization.\ Interestingly, this change in the behavior of the eigenvectors takes place at the onset of the small-world transition possibly arising due to the fact that for $p_{r}$ $\leq$ $0.01$, there exists a decrease in the characteristics path length ($r$) co-existing with a high
clustering coefficient (CC = 3/4).\
It is well known that a higher clustering drives localization of the eigenvectors.\ On the other hand, for $p_{r}$ $\geq$ $0.01$, there exists a significant decrease in CC with $r$ being small, eigenvectors undergo continuous decrease in the degree of localization with an increase in randomness in connections for $p_{r}$ $\geq$ $0.01$.\ 
We would also like to emphasize here that distorting the initial regular network
topology by rewiring few connections does not lead to localization of the eigenvectors, instead, it drives them toward the critical states with $0.4<D_{2}<0.90$.\ Further, it requires a  very few  rewiring,  i.e., a small amount of randomness from the regular structure
to achieve the critical states which we have captured here using the IPR statistics.\ The probability density function of
logarithm of IPR remains the scale-invariant for the critical rewiring probability corresponding to the transition.\ 
Our work be useful to understand various dynamical processes occurring on the small-world networks. For instance,  in \cite{epilepsy}, epilepsy in small-world neural networks was investigated and it was argued that network activities depend on the proportion of long-distance connections. For this particular exmaple, for small, intermediate and high proportion of long-distance connections, the network activity was shown to behave as normal, seizure and bursts, respectively. Normal activity was characterized by low population of firing rates neurons. The Seizure activity was characterized  by  a  significantly  higher  population firing rates while burst activity in the network was characterized by higher firing rates which rise and falls rapidly. A spontaneous active potential in one neuron was shown to lead to the activity in neurons having common post postsynaptic target. Thus, once a wave got initiated, it could give rise to new waves of activity in other regions through the long-distance connections. In this paper, we have shown that for small proportion of long distance connections ($p_{r}$ $<$ 0.01), eigenvectors are more localized as compared to those for higher $p_{r}$ values. Thus, it suggests that probability that if a wave has been initiated will generate another wave through long-distance connections is less since it dies out at the local region perhaps due to constructive interference making this region to behave normal. On the other hand, for intermediate proportion of long distance connections ( 0.01 $\leq$ $p_{r}$ $<$ 0.1), the eigenvectors are less localized as compared to those at the small rewiring probability thus there is a finite probability that if a wave is initiated can initiate new waves through the long-distance connections which may lead to seizure. Finally, at higher rewiring probability eigenvectors are again least localized which can lead to burst activity in network.

\section{Acknowledgments}
 S.J. acknowledges Govt of India, BRNS  Grant  No.  37(3)/14/11/2018-BRNS/37131 for financial support.

\bibliographystyle{elsarticle-num-names}
\bibliography{sample.bib}

\end{document}